\documentclass[11pt]{article}
\usepackage{graphicx}  

\newcommand{\al}{\alpha} 
 
\newcommand{\als}{\frac{\alpha_s}{\pi}} 
\newcommand{\alsb}{\left(\frac{\alpha_s}{\pi}\right)} 
\newcommand{\mts}{M_\tau^2}

\newcommand{\ice}[1]{\relax}
\newcommand{\be}{\begin{equation}}
\newcommand{\ee}{\end{equation}}
\newcommand{\ba}{\begin{eqnarray}}
\newcommand{\ea}{\end{eqnarray}}
\newcommand{\nn}{\nonumber}
\newcommand{\MSsch}{{\overline{\rm MS}}}
\newcommand{\g}{\gamma}
\begin{document}
\begin{flushright}
MZ-TH/02-26\\
November 2002\\
\end{flushright}
\vspace*{0.4cm}

\begin{center}
{\large \bf Extraction of the strong coupling constant 
and\\ 
strange quark mass from semileptonic $\tau$
decays\footnote{Lecture given at International School ``Heavy Quark
Physics'' Dubna, May 28 - June 5, 2002 }
}
\vskip 1cm
{\bf A.~A. Pivovarov}

Institut f\"ur Physik der Johannes-Gutenberg-Universit\"at,\\
Staudinger Weg 7, 55099 Mainz, Germany

and 

Institute for Nuclear Research of the 
Russian Academy of Science,\\ 
117312 Moscow, Russia

\vskip 0.5cm
{\bf Abstract}
\end{center}
\vskip -0.3cm
In this lecture I present a pedagogical introduction to
the low-energy phenomenology of light flavors.
The renormalization 
scheme freedom in defining QCD parameters is
discussed.
It is shown in some details how one can extract 
an accurate numerical value for the strong coupling constant 
from the $\tau$-lepton decay rate into hadrons.
As a related topic I discuss some peculiarities of 
definition of the quark mass 
in theories with confinement and describe the strange quark mass 
determination from data on $\tau$-lepton decays 
employing contour resummation 
which is a modern technique 
of the precision analysis in perturbative QCD.

\section{Definition of QCD parameters $\al_s$ and $m_s$}
Quantum ChromoDynamics (QCD) 
as a field theory for describing strong interactions
is given by the Lagrangian
\[
{\cal L}_{\rm QCD}= \sum_{f}  {\bar q_f}(i\g_\mu \partial^\mu
+g_s \g^\mu G_\mu^a t^a - m_f)q_f 
-\frac{1}{4}G^a_{\mu\nu}G^a_{\mu\nu}
\]
where $G_\mu^a$ is a non-Abelian gluonic field and 
$G_{\mu\nu}^a$ is the field strength tensor~\cite{qcd}.
There are six quark flavors $q=(u,d,s,c,b,t)$, three of which 
($u,d,s$) are called light while the other three are heavy.
There is a close analogy with QED -- the Abelian gauge theory 
for describing the electromagnetic interaction of charged leptons.
The QED Lagrangian for charged leptons reads
\[
{\cal L}_{\rm QED}= \sum_{l}  {\bar l}(i\g_\mu \partial^\mu+e \g^\mu
A_\mu - m_l)l -\frac{1}{4}F_{\mu\nu}F_{\mu\nu}
\]
where $F_{\mu\nu}$ is the electromagnetic field strength tensor.
In the standard model 
there are three charged leptons $l=(e,\mu,\tau)$: electron and muon are very
light at the hadronic mass scale of order $1~{\rm GeV}$
while the $\tau$ lepton is rather heavy with  
a mass $M_\tau = 1.777~{\rm GeV}$~\cite{PDG}.

The interaction is given by a vertex in the Lagrangian 
and normalized to a coupling constant $g_s$
at a tree level of perturbation theory (PT).
A full theory (beyond the tree level) introduces a dressed vertex 
that eventually determines a coupling constant after renormalization.
For the renormalizable models of quantum field theory as QED and QCD
the PT dressing is straightforward and can in principle be
done at any finite order of the expansion in the coupling constant. 
However, the dressing procedure is not unique because renormalization of loops
can introduce some freedom in the choice of
finite parameters through a particular definition of 
counterterms~\cite{bogshirk}.

Leptons can be detected as asymptotic states in the scattering
processes that allows
one to relate the coupling -- for instance, 
the fine structure constant $\al$ -- 
and the particle mass $m_l$ to observables very directly.
Thus, the electron-photon scattering at low energies can be used 
to define the coupling constant (the fine structure constant, in fact) 
$\al=e^2/4\pi$ through the Thompson cross section.
This is a natural definition directly through a physical observable.
Lepton mass is also directly related to experiment:
it can be defined as a physical mass of the asymptotic state
(a position of the pole of the lepton propagator). 
In QCD there is a phenomenon of 
confinement and no asymptotic states of quarks and/or
gluons can be observed. Only the colorless 
hadrons appear as the asymptotic states. 
Therefore, definitions of the coupling constant and quark masses in QCD 
are less direct than in QED. At the theoretical level of the given Lagrangian 
they are very similar though: a vertex for the coupling 
and the propagator for a mass.
To determine numerical values for the coupling and quark masses 
one should turn to experiment.
As there is no possibility to measure these quantities directly
one should specify the research area as the definitions of the
parameters can be adopted to specific experiments (basically to a
corresponding energy scale). 
In this lecture I will talk about
$\tau$-lepton physics which is the area of low-energy hadron
phenomenology: hadronic states have an energy $E<M_\tau$.
The particles which can be observed, for instance, in the process 
of $e^+e^-$ annihilation are
$\rho$, $\omega$, $\varphi$, $\pi\pi$.
In $\tau$ decays $\tau\to \nu + hadrons$ 
one can in addition see $\pi$, $a_1(1260)$, $K$, $K^*(892)$.

The primary difficulty for extracting the QCD coupling 
from experimental data is that the $qqg$ vertex cannot be ``directly'' measured.
Indeed, a typical process (vertex) with hadrons at low energies is 
$\rho\to \pi\pi$ which is not directly expressed theoretically 
through the quark-gluon interaction vertex. 
Therefore, extracting $\al_s$ (defined in terms of quark-gluon vertex
in the Lagrangian) from the experimental quantity as, for instance, 
the $\rho$-meson decay width $\Gamma(\rho\to\pi\pi)$ is highly nontrivial.
One can also see a propagation of the pion
but not that of a quark. No mass shell for the quark (or gluon) is 
seen in the experiment.
Thus, in QCD there is no preferable definition of parameters 
related to experiment. 
Then the only guidance for the choice of a particular definition 
of the QCD parameters
in PT is the technical convenience (and also some general requirements 
such as gauge invariance)~\cite{Bigi:1994em}.

Presently dimensional regularization is overwhelmingly used 
in many loop calculations. 
The renormalization procedure is usually a minimal one -- subtraction of poles
in $\varepsilon= (D-4)/2$ where $D$ is space-time dimension
(an arbitrary complex number formally introduced for the 
regularization purposes). 
This procedure is quite abstract and 
remote from experimental quantities. 
Thus, formally, the $\MSsch$-scheme coupling constant is defined by  
$\al_s^{\MSsch}(\mu) = Z_\al^{\MSsch}(\mu) \al_s^B$ where 
$\al_s^B$ is a bare coupling constant.
This definition of the renormalized coupling constant is not unique. 
In the momentum subtraction (MOM) scheme
$\al_s^{\rm MOM}(\mu)\sim \Gamma^{\rm ren}_{\rm qqg}
(p_1^2=p_2^2=p_3^2=\mu^2)$.
Since the definition is not unique 
the renormalization scheme freedom emerges. 
It is controlled by the renormalization group and can be
conveniently parameterized by the coefficients $\beta_{2,3,\ldots}$ 
of the $\beta$-function~\cite{rgclass,stevenson}
\[
\mu\frac{\partial}{\partial \mu}\al_s(\mu)
=2\beta(\al_s), \quad 
\beta(\al_s)=-\beta_0 \al_s^2-\beta_1 \al_s^3-\beta_2 \al_s^4
-\beta_3 \al_s^5+O(\al_s^6)
\]
and, therefore, any particularly defined coupling constant 
$\al_s^{sch}$ generates an associated
$\beta^{sch}(\al_s^{sch})$-function.
The same is true for the definition of the quark mass. In 
the $\MSsch$-scheme one defines the renormalized mass as 
$m^{\MSsch}(\mu) = Z_m^{\MSsch}(\mu) m^B$ 
with an associated $\gamma$-function
$(\mu\partial/\partial \mu) m(\mu)=2\gamma(\al_s)m(\mu)$.
Functions $\beta(\al_s)$ and $\gamma(\al_s)$ are known 
up to four-loop approximation in PT~\cite{beta3ref}.
The pole mass is difficult to define for light quarks since numerically 
even the strange quark is very light $m_s\sim \Lambda_{\rm QCD}$.

In fact, a QCD coupling constant can also be defined beyond PT 
through physical observables.
For light flavor phenomenology it can directly be defined 
through the cross section of $e^+e^-$ annihilation
\[
1+\al_s(s) \sim \frac{\sigma(e^+e^-\to {\rm hadrons})}
{\sigma(e^+e^-\to \mu \bar \mu)}
\]
while for heavy quark physics
the definition based on the heavy-quark static potential
$\al_V({\vec q}^2)\sim V_{q\bar q}(\vec q^2)$
can be useful~\cite{physcharge}.

\section{Kinematics of semileptonic $\tau$ decays}
The differential decay rate of the $\tau$ lepton into
an hadronic system  $H(s)$ with a total squared energy $s$
\[
\frac{d\sigma(\tau\to \nu H(s))}{ds}\sim \left(1-\frac{s}{M_\tau^2}\right)^2
\left(1+\frac{2s}{M_\tau^2}\right)\rho(s)
\]
is determined by the hadronic spectral density $\rho(s)$
defined through the correlator of weak currents. For the
$(ud)$ current 
$j_{\mu}^W(x) = \bar{u}\gamma_{\mu}(1-\gamma_5) d$
one finds
\be
\label{corr}
i\!\int\!\! \langle Tj_{\mu}^W(x)j_{\nu}^{W+}(0) \rangle e^{iqx}dx
=(q_\mu q_\nu - q^2 g_{\mu\nu})\Pi^{\rm had}(q^2), 
\Pi^{\rm had}(q^2)=\int \frac{\rho(s)ds}{s-q^2}
\ee
with $\rho(s)\sim {\rm Im}~\Pi^{\rm had}(s+i0)$, $s=q^2$.
The function $\Pi^{\rm had}(Q^2)$ with $Q^2=-q^2$
is calculable in pQCD far from the physical cut
as a series in the running coupling constant $\al_s(Q^2)$.

NonPT effects (power corrections) are included using
OPE for the correlator~(\ref{corr})
at small distances as $x\to 0$ in Euclidean domain
(that corresponds to large $Q^2$)
through phenomenological characteristics of the vacuum such as
gluon and quark condensates~\cite{SVZ}.
The lattice approximation for the evaluation of 
the correlator $\Pi^{\rm had}(Q^2)$ beyond PT can also be 
used~\cite{lattice}.
This is a basis for theoretical description of semileptonic 
$\tau$ decays in QCD.

Integrating the function $\Pi^{\rm had}(z)$ 
over a contour in the complex $q^2$ plane beyond the physical cut
$s>0$ one finds that 
for particular weight functions some 
integrals of the hadronic spectral density $\rho(s)$
can be reliably computed in PT~\cite{cont1,cont2}.
Indeed, due to Cauchy theorem one gets
\[
\oint_C \Pi(z)dz = \int_{\rm cut} \rho(s) ds\ .
\]
Using the approximation
$\Pi^{\rm had}(z)|_{z\in C} \approx \Pi^{\rm PT}(z)|_{z\in C}$
which is well justified sufficiently far from the physical cut 
one obtains
\[
\oint_C \Pi^{\rm had}(z)dz = \int_{\rm cut} \rho(s) ds 
= \oint_C \Pi^{\rm PT}(z)dz 
\]
i.e. the integral over the hadronic spectrum can be evaluated in pQCD.
The total decay rate of the $\tau$ lepton 
written in the form of an integral along the cut
\[ 
R_{\tau S=0}=\frac{\Gamma(\tau \rightarrow H_{S=0} \nu)}
{\Gamma(\tau \rightarrow l \bar{\nu} \nu )}
\sim \int_{\rm cut}\left(1-\frac{s}{M_\tau^2}\right)^2
\left(1+\frac{2s}{M_\tau^2}\right)\rho(s)ds
\]
is precisely the quantity that one can reliably 
compute in pQCD~\cite{tauanal}.

\section{PT analysis in QCD}
For technical reasons (no overall UV divergence)
a derivative of the correlator $\Pi^{\rm had}(Q^2)$
is often used for presenting results of PT evaluation
\[
D(Q^2)=-Q^2\frac{d}{d Q^2}\Pi^{\rm had}(Q^2)
=Q^2\int \frac{\rho(s)ds}{(s+Q^2)^2}\, .
\]
The PT expansion for the $D$-function in terms of $a_s(Q)=\al_s(Q)/\pi$ reads
$D(Q^2)=1+a_s(Q)+k_1 a_s(Q)^2 +k_2 a_s(Q)^3+ k_3 a_s(Q)^4$.
In the $\overline{\rm MS}$-scheme 
\[
k_1=\frac{299}{24} - 9\zeta(3),\quad
k_2=\frac{58057}{288} - \frac{779}{4}\zeta(3) +
\frac{75}{2}\zeta(5)
\]
with $\zeta$-function equal $\zeta(3)=1.202...$ and $\zeta(5)=1.037...$
The above numerical values for $k_{1,2}$
summarize the results of three and four
loop PT calculations~\cite{eek20}. Numerically,
one finds
$D(Q^2)=1+a_s+1.64 a_s^2 + 6.37 a_s^3 + k_3 a_s^4$.
Coefficient $k_3$ is known only partly~\cite{Baikov:2001aa}.
It is retained to obtain a feeling for the possible 
magnitude of the $O(\al_s^4)$ correction.
The decay rate of the $\tau$ lepton 
into nonstrange hadrons is written in the form
\[
R_{\tau S=0}=\frac{\Gamma(\tau \rightarrow H_{S=0} \nu)}
{\Gamma(\tau \rightarrow l \bar{\nu} \nu )}
= 3 |V_{ud}|^2 (1+\delta_P + \delta_{NP})\ .
\]
Here the first term is the parton model 
result, the second term $\delta_P$ represents pQCD effects. 
NonPT effects are small, $\delta_{NP}\approx 0$,
in the factorization approximation for
the four-quark vacuum condensates which 
is quite accurate~\cite{tauanal,nonpttau}.

The experimental result
$R_{\tau S=0 }^{\rm exp}=3.492 \pm 0.016$
leads to $\delta_{P}^{\rm exp}=0.203\pm0.007$~\cite{exp12}.
In the $\overline{\rm MS}$-scheme the correction 
$\delta_{P}$ is given by the series
\[
\delta_P^{\rm th} = a_s + 5.2023 a_s^2
+26.366a_s^3 +(78.003+k_3)a_s^4 + O(a_s^5)
\]
with $a_s$ taken at the scale 
$\mu=M_\tau$.
Usually one extracts a numerical value for 
$\al_s(M_\tau)$
by treating the first three terms of the expression
as an exact function -- the cubic polynomial
$a_s + 5.2023 a_s^2+26.366a_s^3=\delta_P^{\rm exp}$.
The solution reads
$\pi a_s^{st}(M_\tau)\equiv \al_s^{st}(M_\tau)
= 0.3404\pm 0.0073_{exp}$.
The error is due to the error of the input experimental value 
$\delta_P^{\rm exp}$.
It is difficult to estimate the theoretical uncertainty of the
approximation for the (asymptotic)
series given by the cubic polynomial (higher order terms).
One criterion is the pattern of convergence of the series 
\[
{ \delta_P^{\rm exp}=0.203=0.108+0.061+0.034+\ldots}
\]
The corrections provide a 100\% change of the leading term.
Another criterion is the order-by-order behavior of the extracted
numerical value for the coupling constant. 
In consecutive orders of PT
\[ 
\al_s^{st}(M_\tau)_{LO}=0.6377,\quad 
\al_s^{st}(M_\tau)_{NLO}=0.3882,\quad
\al_s^{st}(M_\tau)_{NNLO}=0.3404 
\]
that translates into a series for the 
coupling constant
\[
\al_s^{st}(M_\tau)_{NNLO}=0.6377-0.2495-0.0478-\ldots
\]
One can take a half(??) of the last term
as an estimate of the theoretical 
uncertainty. It is only an indicative estimate.
No rigorous justification can be given for such an assumption
about the accuracy of the approximation without knowledge
of the structure of the whole series.
The uncertainty obtained in such a way
$\Delta \al_s^{st}(M_\tau)_{th}=0.0478/2=0.0239\gg
0.0073_{exp}$
is much larger than that from experiment.
This is a challenge for the theory: the accuracy of 
theoretical formulae cannot
compete with experimental precision. 
Assuming this theoretical uncertainty one has
\[
\al_s^{st}(M_\tau)_{NNLO}=0.3404\pm 0.0239_{th}\pm
0.0073_{exp}
\]
Theory dominates the error. Still it is not the whole story.
Now one can choose a different expansion 
parameter. The simplest way is to change the
scale of the coupling along the RG trajectory 
$M_\tau\to 1~{\rm GeV}$. 
In terms of $a_s(1~{\rm GeV})$ one finds a series
$a_s(1) +  2.615 a_s(1)^2
+1.54 a_s(1)^3=\delta_P^{\rm exp}$.
The solution for the coupling constant is $\al_s(1~{\rm GeV})=0.453$.
The convergence pattern for the correction 
$\delta_P^{\rm exp}$ is
\[
\delta_P^{\rm exp}=0.203 = 0.144 + 0.054 + 0.005
\]
and for the numerical value of the coupling constant
\[
\al_s(1)= 0.453=0.638-0.177-0.008\, .
\]
Should one conclude that now the accuracy is much better?
What would be an invariant criterion for the precision 
of theoretical predictions obtained from PT, i.e. finite number 
of terms of asymptotic series?
Thus, one sees that the renormalization scheme dependence can 
strongly obscure the heuristic evaluation of the accuracy of
theoretical formulae in the absence of any information on the
structure of the whole PT series. The final result for the standard 
reference value of the 
coupling constant normalized to the $Z$ boson mass $M_Z$ 
reads~\cite{krajal}
\[
\al_s(M_Z)_{\tau}= 0.1184\pm 0.0007_{exp}\pm 0.0006_{hq~mass}
\pm 0.0010_{th=truncation}
\]
with a theoretical uncertainty that mainly comes from the truncation 
of the PT series. The world average value given by the 
Particle Data Group reads~\cite{PDG}
\[
\al_s(M_Z)_{\rm PDG}^{\rm av}= 0.1172\pm 0.002\, .
\]

To reduce a renormalization scheme dependence of the theoretical
analysis one 
should use several observables simultaneously~\cite{several}.
Such a possibility has recently emerged in study of 
$\tau$ decays since experimental
data on Cabibbo suppressed ($S=1$) channel appeared~\cite{msNPBPP}.
For strange hadrons $H_{S = 1}$ 
($us$ part of the weak current: Cabibbo suppressed decays)
the decay rate becomes
\[ 
R_{\tau S=1}=\frac{\Gamma(\tau \rightarrow H_{S=1} \nu)}
{\Gamma(\tau \rightarrow l \bar{\nu} \nu )} 
= 3 |V_{us}|^2 (1+\delta_P^\prime + \delta_{NP}^\prime)\ .
\]
The first term (``{1}'') is the parton model result, 
the second term $\delta_P^\prime$ gives pQCD effects. 
Small $s$-quark mass effects for Cabibbo suppressed part of the rate are 
taken in PT at the leading order in the ratio $m_s^2/\mts$
since $m_s\ll M_\tau$
\[
\delta_P^\prime(\al_s,m_s)
=\delta_P(\al_s)+\frac{m_s^2}{\mts}\Delta_m(\al_s)
\]
with $\delta_P(\al_s)$
being a correction in massless approximation
for light quarks that is well justified for nonstrange decays 
($ud$ part)
since $u,d$ quarks are very light indeed
$m_u+m_d=14~{\rm MeV}$~\cite{lightmass1,lightmass2}.
The correlator of the weak charged strange current 
$j_{\mu}(x) = \bar{u}\gamma_{\mu}(1-\gamma_5) s$ with 
a finite $s$-quark mass is not transverse 
\[
i\int dx e^{iqx}
\langle T j_{\mu}(x) j_{\nu}^{\dagger} (0) \rangle
= q_{\mu}q_{\nu} \Pi_q(q^2)+g_{\mu\nu}  \Pi_g(q^2)\, .
\]
Retaining the first order term of expansion in the small ratio
$m_s^2/q^2$ one finds the $m_s^2$ correction to the invariant
functions $\Pi_{q,g}(q^2)$
\[
\Pi_q(q^2)=\Pi(q^2)+3\frac{m_s^2}{q^2}\Pi_{mq}(q^2),\quad
\Pi_{g}(q^2)=-q^2\Pi(q^2)+\frac{3}{2} m_s^2\Pi_{mg}(q^2)
\]
where $\Pi(q^2)$
is an invariant function for the mass zero case. 
The functions $\Pi_{q,g}(Q^2)$ are computable in
QCD perturbation theory within operator product expansion 
for $Q^2\rightarrow \infty$. Thus, the experimental data on 
$\tau$ lepton decays are
theoretically described by three independent invariant functions (form
factors) which can be analyzed simultaneously that may help
to reduce uncertainties introduced by the renormalization scheme
freedom.

In the actual analysis one can factor out the renormalization scheme
freedom to large extent by introducing an effective scheme with 
definitions of effective quantities 
$a$, $m_q^2$, $m_g^2$ through the relations~\cite{krajms}
\begin{eqnarray} 
\label{effdef}
\qquad \qquad \quad -Q^2\frac{d}{dQ^2}\Pi(Q^2)
&=&1 + a(Q^2)\, ,   \nn \\
- m_s^2(\mts) Q^2\frac{d}{dQ^2}\Pi_{mg}(Q^2)
&=& m_g^2(\mts)C_g(Q^2)\, , \\
\qquad \qquad m_s^2(\mts) \Pi_{mq}(Q^2)
&=& m_q^2(\mts)C_q(Q^2) \nn \, .
\end{eqnarray}
Here $C_{q,g}(Q^2)$ are coefficient functions of mass corrections.
They are conveniently normalized by the requirement $C_{q,g}(\mts)=1$.
In terms of the $\MSsch$ scheme quantities 
$\al_s\equiv \al_s(\mts)$ and $m_s\equiv m_s(\mts)$
the effective parameters in eq.~(\ref{effdef}) read
\ba\label{effparam}
a(\mts)&=& 
\als+k_1\alsb^2 + k_2\alsb^3 + k_3\alsb^4 + {\cal O}(\al_s^5)  \, , \nn\\  
\label{effmg}
m_g^2(\mts)&=& 
m_s^2(\mts)(1+\frac{5}{3}\als +k_{g1}\alsb^2 +k_{g2}\alsb^3 
+ {\cal O}(\al_s^4))\, ,\nn \\
\label{effmq}
m_q^2(\mts)&=& 
m_s^2(\mts)(1+\frac{7}{3}\als + k_{q1}\alsb^2+ k_{q2}\alsb^3
+ {\cal O}(\al_s^4))\, \nn.
\ea
Numerical values for the coefficients $k_3$, $k_{q2}$
are unknown though their estimates within various intuitive approaches
can be found in the literature~\cite{highorder}.

For confronting the experimental data obtained in $\tau$ decays with theory
of strong interactions one uses special integrals of 
the hadronic spectral density $\rho(s)$ (called spectral moments)
of the form 
\[
{\cal M}_{kl}^{\rm exp}
=\int_0^{M_\tau^2} \rho(s) \left(1-\frac{s}{M_\tau^2}\right)^k  
\left(\frac{s}{M_\tau^2}\right)^l \frac{ds}{M_\tau^2}
\]
which is suitable for experiment as the spectrum 
$\rho(s)$ is measured.
Another representation of the moments is suitable for 
the perturbation theory 
evaluation in QCD. Due to analytic properties of the functions 
$\Pi_{q,g}(Q^2)$
the moments can be rewritten as contour integrals in the complex $q^2$
plane:
\[
{\cal M}_{kl}^{\rm th} = \frac{i}{2\pi}
\oint_C \Pi_\#(z)\left(1-\frac{z}{\mts}\right)^k   
\left(\frac{z}{\mts}\right)^l 
\frac{dz}{\mts}\, .
\]
Technically one chooses a circular contour
in the complex $Q^2$-plane
with \mbox{$Q^2=\mts e^{i \phi}$}, \mbox{$-\pi< \phi<\pi$}
that converts all invariant amplitudes $\Pi_{q,g}(Q^2)$
to the certain functions of the angle $\phi$.
The evolution of the functions
$\Pi_{q,g}(Q^2)$ along the contour in the complex plane 
is governed by the renormalization group~\cite{Pivtau}.
For the massless case corresponding to the analysis of data in Cabibbo
favored channel the only relevant quantity is
the running ``coupling constant'' 
$a(Q^2)\to a(\phi)$
for \mbox{$Q^2=\mts e^{i \phi}$} that serves as the expansion parameter
for the function $\Pi(Q^2)$ along the contour. 
Note that there are no higher order corrections in the effective
scheme by definition.
The renormalization group 
evolution is
determined by the effective $\beta$ function 
for the effective coupling constant~\cite{grunberg,effsch} 
\[ 
 - i \frac{d}{d  \phi} a(\phi) = \beta(a(\phi))\, , 
\quad a(\phi = 0) = a(\mts)\, ,
\]
and the anomalous dimension for the running mass
\[
- i\frac{d}{d  \phi} m_s(\phi) 
= \gamma_m[a(\phi)] m_s(\phi) \, , \quad m_s(\phi = 0) = m_s\, .
\]
The initial values for $a(\phi)$ and 
$m_s(\phi)$ are extracted from fit to data. 
Thus, we have three quantities $a(Q)$, $C_q(Q)$, $C_g(Q)$ 
associated with three invariant functions
$\Pi(Q^2)$, $\Pi_{q,g}(Q^2)$ describing the $\tau$ system
in the considered approximation.

The RG equations for the set of quantities
$\{a(Q),C_q(Q),C_g(Q)\}$ are
\[
Q^2\frac{d}{dQ^2} a(Q^2)=\beta(a),\quad
Q^2\frac{d}{dQ^2}C_{g,q}(Q^2)=2 \gamma_{g,q}(a)C_{g,q}(Q^2)\, .
\]

\begin{figure}[t]
\label{fig1b}
 \includegraphics[width=.45\textwidth]{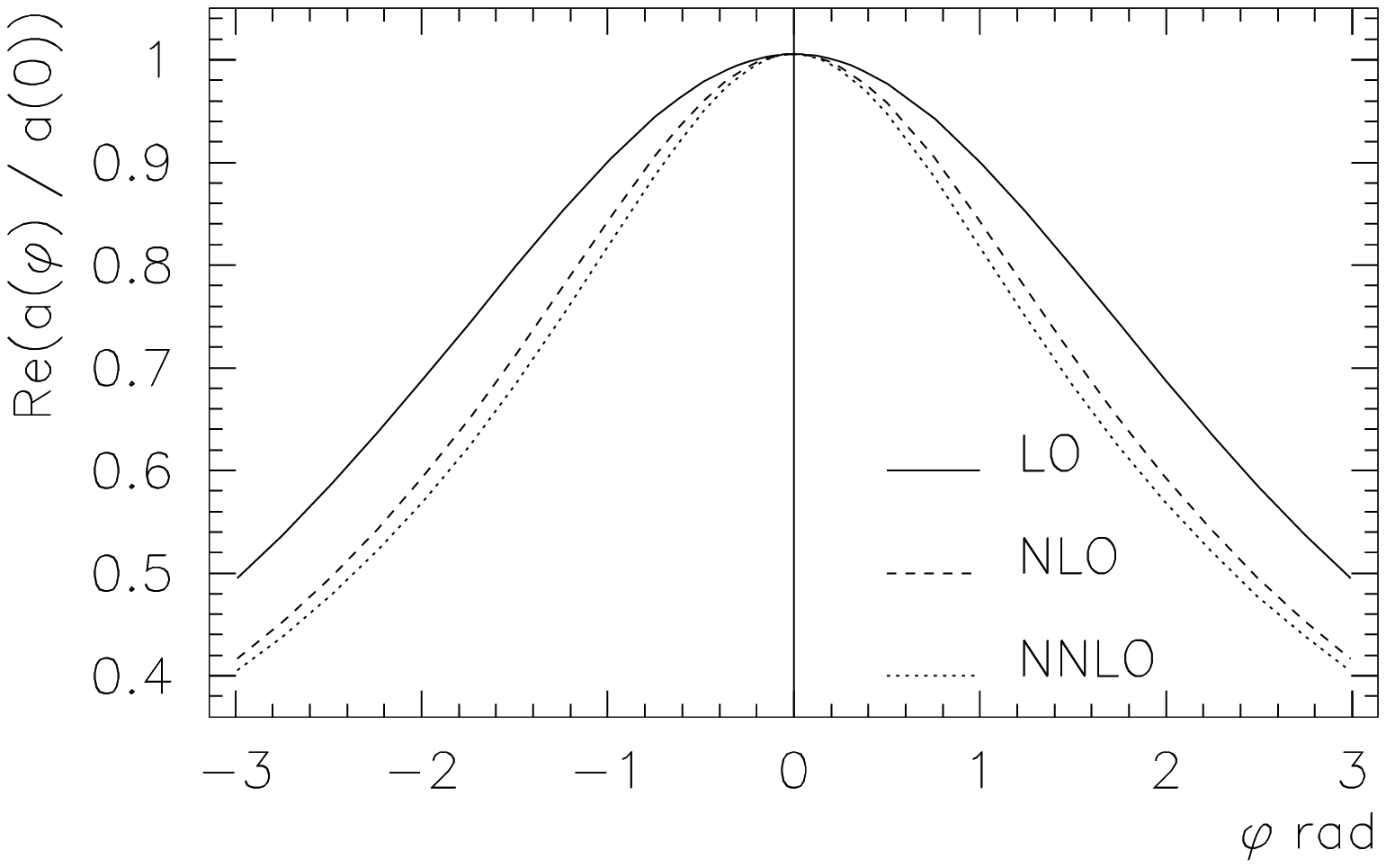}
       \quad \quad\quad
\includegraphics[width=.45\textwidth]{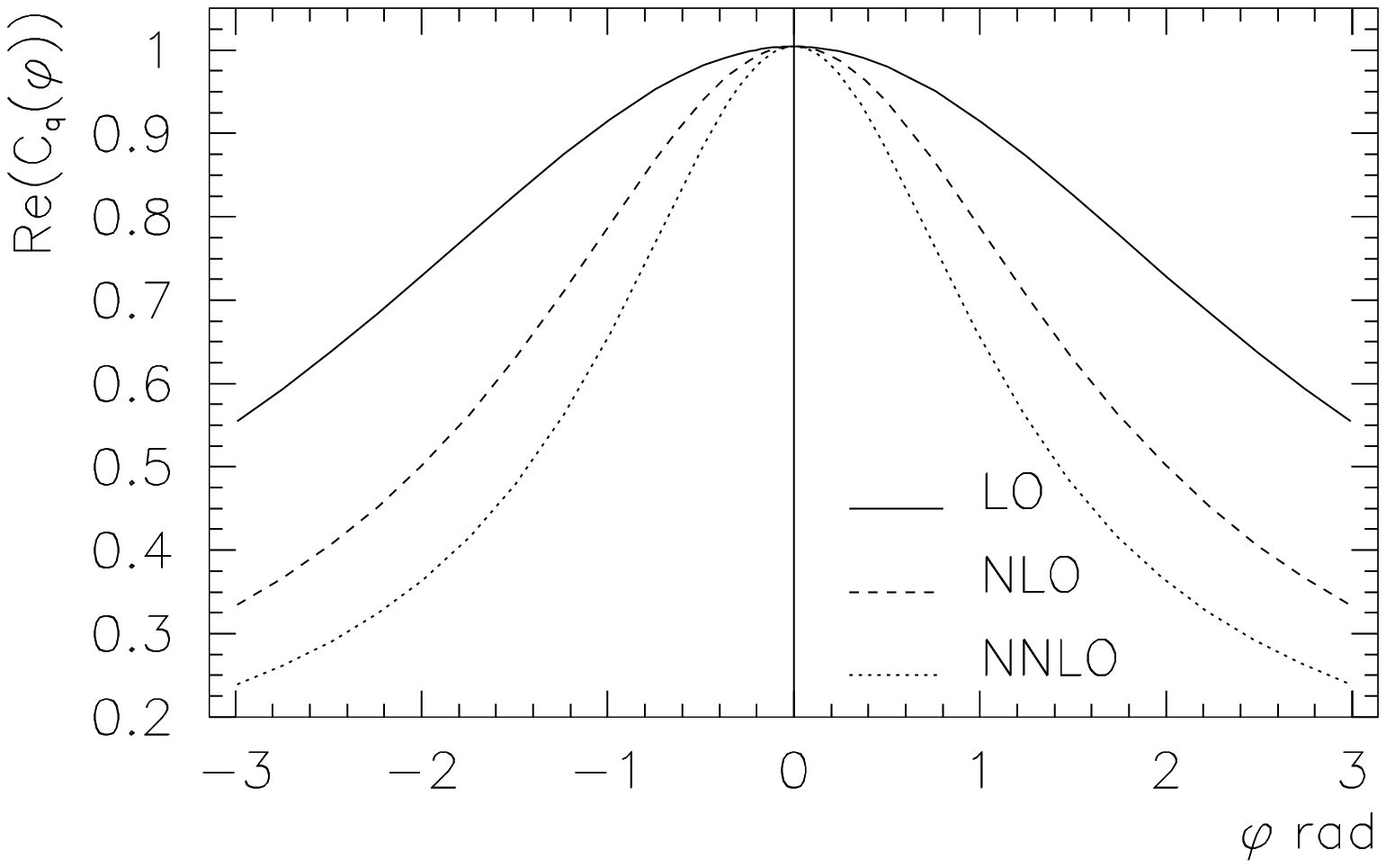}
       \caption{\label{runc}
Running of the functions $a(\phi)$ and $C_q(\phi)$
on a circular contour 
in the complex plane calculated at LO, NLO and NNLO 
(left:  $a(\phi)$; right: $C_q(\phi)$; real parts only)}
\end{figure}

The RG functions $\beta(a)$ and $\gamma_{g,q}(a)$
are given by the expressions~\cite{krajms} 
\[
\frac{-4\beta(a)}{9 a^2}=1 + 1.778a 
+ 5.24 a^2 + a^3( -34 + 2 k_3)\, ;
\]
\[
\frac{-\gamma_g(a)}{a}=1 + 4.03 a + 17.45 a^2  
+ a^3( 249.59 - k_3 )\, ;
\]
\[
\frac{-\gamma_q(a)}{a}=1 + 4.78 a + 32.99 a^2 
+ a^3(-252 - k_3 + 3.4 k_{q2})\, .
\]
The solution of the renormalization group equation for the 
effective coupling constant $a(\phi)$ converges well
when the higher order corrections of the $\beta$-function
are included. The change from the next-to-leading order (NLO) solution
to the next-to-next-to-leading order (NNLO)
is small. The behavior of
the coefficient function $C_g(\phi)$ (not shown) 
related to the contributions of
spin one particles is rather similar to that 
of the coupling constant.
However, the convergence pattern of the function $C_q(\phi)$ is much 
worse (see Fig.~1). It seems that the $\gamma_q$-function 
has already shown up an asymptotic growth in the 
next-to-next-to-leading order 
which will limit the precision of our results.
Note that the fact that the function $C_q(\phi)$
can behave wilder in higher order of perturbation theory is expected 
since this function is more infrared sensitive than the coupling
constant $a(\phi)$ and the function $C_g(\phi)$.

\begin{table}[t]
\caption{\label{T1}Coefficients of eq.~(\ref{finaleq})}
\[ 
\begin{array}{||c||c|c|c||c|c|c||}
\hline
(k,l) & A_{kl}^{\rm{LO}} & A_{kl}^{\rm{NLO}}  
& A_{kl}^{\rm{NNLO}} & B_{kl}^{\rm{LO}} & B_{kl}^{\rm{NLO}}  
& B_{kl}^{\rm{NNLO}}  \\ 
\hline
(0,0) & 1.361  & 1.445  & 1.434  &0.523 & 0.601  & 0.625 \\
(1,0) & 1.568  & 1.843  & 1.976  &0.441 & 0.552  & 0.601 \\
(2,0) & 1.762  & 2.282  & 2.646  &0.390 & 0.530  & 0.607 \\ 
\hline 
\end{array}
\]
\end{table}

The numerical value for the $s$-quark mass $m_s$
is extracted from the difference between moments of 
Cabibbo-favored ($ud$-type)
and and Cabibbo-suppressed ($us$-type) decay rates
\[
\delta \! R_\tau^{kl} 
=\frac{R_{\tau S=0}^{kl}}{|V_{ud}|^2} 
-\frac{R^{kl}_{\tau S=1}}{|V_{us}|^2}, \quad
R_{\tau S=0,1}^{kl} 
= \int_0^{\mts} \!\!ds \left(1- \frac{s}{\mts} \right)^k
\left( \frac{s}{\mts} \right)^l 
\frac{d R_{\tau S = 0,1}}{ds}\, .
\]
The theoretical expression for the $m_s^2$
corrections to the moments $(k,l)$ corresponding to the above
experimental quantity
is given by the contour integral in the complex $q^2$ plane
\[
\frac{3i m_s^2}{\pi} 
\oint_C \left( 1- \frac{z}{\mts} \right)^{2+k} 
\left(\frac{z}{\mts} \right)^l \left(\frac{\Pi_{mq}(z)}{z} 
- \frac{\Pi_{mg}(z)}{\mts}\right)\frac{dz}{\mts}\, .
\]
In the theoretical expression for the 
difference $\delta \! R_\tau^{kl}$ we neglect terms of 
the order $m_s^3/M_\tau^3$, set the $u$- and $d$-quark masses to zero, 
and retain only the most important term linear in $m_s$.
Within operator product expansion 
the coefficient of this term is given by the quark
condensate. The final result for the difference reads~\cite{krajms}
\be
\label{finaleq}
\delta \! R_\tau^{kl} = 3 S_{EW} 
\left(6\frac{m_s^2}{M_\tau^2} (\omega_q A_{kl}+\omega_g B_{kl})
- 4\pi^2\frac{m_s}{M_\tau}\frac{\langle \bar s s \rangle}{M_\tau^3} 
T_{kl} \right)
\ee
with $S_{EW}=1.0194$~\cite{ewcorrsir}.
Here $m_{q,g}^2=\omega_{q,g} m_s^2$ with  
$\omega_q =  1.73 \pm 0.04$, $\omega_g = 1.42 \pm 0.03$.
We use the relation between vacuum 
condensates of strange and nonstrange quarks
\[
\langle \bar s s \rangle 
= (0.8 \pm 0.2 ) \langle \bar u u \rangle
\]
and the numerical value
$\langle \bar u u \rangle = - (0.23~\rm{GeV} )^3$~\cite{gammas}.
In the leading order approximation of the QCD perturbation theory
for the coefficient function of the quark condensate  
the quantities $T_{kl}$ multiplying the quark condensate
are given by the expression
\[
T_{kl} = 2\left( \delta_{l,0} (k+2) - \delta_{l,1} \right)\, .
\]
The numerical values for the first few coefficients $T_{kl}$ 
read
\[
T_{00} = 4, \quad T_{10}  = 6, \quad
T_{20} = 8,  \quad  T_{01} = -2, \quad T_{11} = -2\, .
\]
The numerical values for the coefficients $A_{kl}$ and $B_{kl}$ 
are given in Table 1.

\begin{table}
\caption{\label{T2}Experimental moments and extracted mass}
\[
\begin{array}{|c|c|c|}
\hline
(k,l)&(\delta\! R_\tau^{kl})^{\rm exp}& m_s(\mts)~{\rm MeV}\\ \hline
(0,0)& 0.394 \pm 0.137 & 130\pm \delta_{00}^{\rm th}(=6) \\
(1,0)& 0.383 \pm 0.078 & 111 \pm \delta_{10}^{\rm th}(=?)\\
(2,0)& 0.373 \pm 0.054 &  95 \pm \delta_{20}^{\rm th}(=?)\\ \hline
\end{array}
\]
\end{table}

Using experimental data one extracts $m_s$ as
given in Table 2. Theoretical prediction for the 
moment $(0,0)$ is the most reliable from PT point of view as
$\delta_{20}^{\rm th}(=?)>\delta_{10}^{\rm th}(=?)
>\delta_{00}^{\rm th}=6~{\rm MeV}$.
The final result reads 
\[
m_s(\mts)=130 \pm 27_{\rm{exp}} \pm 3_{\langle\bar{s}s \rangle} 
\pm 6_{\rm{th}}~{\rm MeV}.
\]
Normalization at $1~{\rm GeV}$ gives
\[
m_s(1~{\rm GeV})=176 \pm 37_{\rm{exp}} \pm 4_{\langle\bar{s}s \rangle} 
\pm 9_{\rm{th}}~{\rm MeV}\, .
\]
Only the moment $(0,0)$ is used for the $m_s$ determination 
as the most reliable one from the PT point of view.
The higher order moments with the weight function 
$(1-s/M_\tau^2)^k$ for large $k$ 
have an uncontrollable admixture 
of higher dimension
condensates that makes them strongly nonperturbative
and, therefore, unreliable for applications based on PT 
calculations~\cite{corr2point}.
The contributions of higher dimension
condensates are unknown and from general considerations the errors
$\delta_{20}^{\rm th}(=?)>\delta_{10}^{\rm th}(=?)$
are expected to be 
much larger than $\delta_{00}^{\rm th}=6~{\rm MeV}$.
The value of the strange quark mass obtained by using 
the effective scheme approach as described in the present paper
is in a reasonable agreement with other 
estimates~\cite{others1,others2,others3}. It is a bit larger than the recent 
lattice determination~\cite{latticems}.

To conclude, the experimental information on $\tau$ decays is 
a reliable source for the precision
determination of the numerical values of important QCD parameters
$\al_s$ and $m_s$. 

\section*{Acknowledgements}
I thank K.G.~Chetyrkin, S.~Groote, J.G.~K\"orner, J.H.~K\"uhn for
useful discussions and fruitful collaboration. 
This work is partially supported by the Russian Fund for Basic Research 
under contracts 01-02-16171 and 02-01-00601 and by the Volkswagen grant. 


\end{document}